\documentclass[aps,prd,onecolumn,nofootinbib,preprintnumbers,superscriptaddress,longbibliography,nobibnotes,notitlepage]{revtex4-1}

\usepackage[pdftex,colorlinks]{hyperref}
\usepackage[dvipsnames]{xcolor}
\definecolor{phthaloblue}{rgb}{0.0, 0.06, 0.54}

\usepackage{bm}

\newcommand{\be}{\begin{equation}}
\newcommand{\ee}{\end{equation}}
\newcommand{\nl}{\nonumber \\}
\newcommand{\x}{\chi}

\newcommand{\Ap}{A^\prime}

\newcommand{\Lag}{\mathscr{L}}
\newcommand{\eps}{\epsilon}

\newcommand{\p}{\prime}

\newcommand{\Eq}[1]{Eq.~\ref{eq:#1}}

\newcommand{\Eqs}[2]{Eqs.~\ref{eq:#1} and \ref{eq:#2}} 
\newcommand{\Sec}[1]{Sec.~\ref{sec:#1}}

\newcommand{\Fig}[1]{Fig.~\ref{fig:#1}}

\newcommand\mat[1]{\begin{pmatrix}#1\end{pmatrix}} 
\newcommand{\bebox}{\begin{empheq}[box=\fcolorbox{light-gray}{light-gray}]{align}}
\newcommand{\eebox}{\end{empheq}}

\newcommand{\cm}{\text{cm}}

\newcommand{\eV}{\text{eV}}

\newcommand{\MeV}{\text{MeV}}
\newcommand{\GeV}{\text{GeV}}
\newcommand{\TeV}{\text{TeV}}

\newcommand{\kpc}{\text{kpc}}

\newcommand{\mn}{\mu \nu}

\newcommand{\g}{\gamma}

\newcommand{\Neff}{N_\text{eff}}

\usepackage[american]{babel}
\usepackage{microtype}
\usepackage{appendix}
\usepackage{booktabs}
\usepackage{feynmp-auto}
\usepackage{gensymb}

\usepackage{mathtools}

\usepackage{amsmath,amssymb,amsthm}
\usepackage{graphicx}
\usepackage{slashed}
\usepackage{tabularx}
\usepackage{empheq}
\usepackage{ifthen}
\usepackage{mathrsfs}
\usepackage{comment}
\usepackage{bbold}
\usepackage[shortlabels]{enumitem}

\usepackage{url}
\usepackage{color}

\definecolor{light-gray}{gray}{0.9}
\definecolor{medium-gray}{gray}{0.7}
\definecolor{darkblue}{rgb}{0.0,0.0,0.6}
\definecolor{red}{rgb}{0.9, 0,0}
\definecolor{navy}{rgb}{0.05, 0.05,0.8}
\definecolor{linkcolor}{rgb}{0.0, 0.28, 0.67}
\definecolor{paleblue}{rgb}{0.69, 0.93, 0.93}  

\usepackage[colorlinks]{hyperref}
\hypersetup{
     colorlinks   = true,
     citecolor    = linkcolor,
     linkcolor    = linkcolor,
     urlcolor    = linkcolor
}

\hypersetup{
    colorlinks=MidnightBlue,
    linkcolor=MidnightBlue,
    citecolor=MidnightBlue,
    filecolor=MidnightBlue,
    urlcolor=MidnightBlue,
    }

\usepackage{feynmp}
\usepackage[utf8]{inputenc}
\DeclareUnicodeCharacter{2212}{-}
\usepackage{orcidlink}

\usepackage{tabularx}
\usepackage{array}
\newcommand{\ra}[1]{\renewcommand{\arraystretch}{#1}}
\makeatletter
\def\hlinewd#1{%
\noalign{\ifnum0=`}\fi\hrule \@height #1 \futurelet
\reserved@a\@xhline}
\makeatother

\newcommand{\beq}{\begin{equation}}
\newcommand{\eeq}{\end{equation}}

\newcommand{\cmps}{\cm^3 \ \text{s}^{-1}}
\newcommand{\sigv}{\langle \sigma v \rangle}
\newcommand{\ml}{m_{\x_1}}
\newcommand{\mh}{m_{\x_2}}

\begin{document}
\preprint{FERMILAB-PUB-25-0227-T}

\title{dSphobic Dark Matter}

\author{Asher Berlin\,\orcidlink{0000-0002-1156-1482}}
\email{berlin@fnal.gov}
\affiliation{Theoretical Physics Division, Fermi National Accelerator Laboratory, Batavia, IL, USA}

\author{Joshua W. Foster \,\orcidlink{0000-0002-7399-2608}}
\email{jwfoster@fnal.gov}
\affiliation{Theoretical Physics Division, Fermi National Accelerator Laboratory, Batavia, IL, USA}
\affiliation{Kavli Institute for Cosmological Physics, University of Chicago, Chicago, IL USA}

\author{Dan Hooper\,\orcidlink{0000-0001-7420-9577}}
\email{dwhooper@wisc.edu}
\affiliation{Department of Physics, University of Wisconsin, Madison, WI, USA}
\affiliation{Wisconsin IceCube Particle Astrophysics Center, University of Wisconsin, Madison, WI, USA}

\author{Gordan Krnjaic\,\orcidlink{0000-0001-7420-9577}}
\email{krnjaicg@fnal.gov}
\affiliation{Theoretical Physics Division, Fermi National Accelerator Laboratory, Batavia, IL, USA}
\affiliation{Kavli Institute for Cosmological Physics, University of Chicago, Chicago, IL USA}
\affiliation{Department of Astronomy \& Astrophysics, University of Chicago, Chicago, IL, USA}

\begin{abstract}
\vspace{0.3cm}
\noindent
We present a mechanism that allows thermal relic dark matter to annihilate efficiently in the Galactic Halo and in galaxy clusters, but not in the lower-velocity environments of dwarf spheroidal (dSph) galaxies. We realize this within a complete model in which the dark matter consists of two distinct states separated by a small mass splitting. An indirect detection signal is generated only through the coannihilations of these two states, requiring both to be present. In the halo of the Milky Way, the dark matter particles in the lighter state can be excited into the long-lived heavier state through scattering. Once excited, these heavier particles can coannihilate with those in the lighter state, yielding a gamma-ray signal with little or no suppression. By contrast, the dark matter particles in dwarf galaxies do not possess enough kinetic energy to be excited, thereby suppressing the coannihilation rate and corresponding indirect detection signals from those systems. This framework breaks the predictive relationship that ordinarily exists between these respective gamma-ray signals and complicates our ability to interpret the results of indirect detection searches. 
\end{abstract}

\maketitle

\section{Introduction}

Thermal freeze out is perhaps the most compelling and predictive mechanism for producing dark matter (DM). 
In this scenario, the DM is initially in chemical equilibrium with the Standard Model (SM) radiation bath, but decouples at a characteristic freeze-out temperature, somewhat below its mass (for a review, see Ref.~\cite{Steigman:2012nb}). For DM particles with roughly weak-scale masses and interactions, this mechanism naturally yields the observed DM density and predicts a thermally-averaged annihilation cross section of  $\langle\sigma v \rangle \sim  10^{-26} \, \cmps$~\cite{Steigman:2012nb}. 

This connection between the DM annihilation cross section and its abundance has motivated many efforts to observe these reactions in different astrophysical settings. These strategies include searches for gamma-rays from the Galactic Center~\cite{Fermi-LAT:2015sau,Fermi-LAT:2017opo}, the Galactic Halo~\cite{Fermi-LAT:2012pls,Fermi-LAT:2013thd}, and the satellite (dwarf) galaxies of the Milky Way~\cite{McDaniel:2023bju,DiMauro:2022hue}.
Related searches have also been conducted using positrons~\cite{Bergstrom:2013jra}, antiprotons~\cite{Calore:2022stf,Cuoco:2016eej,Cui:2016ppb,Cholis:2019ejx}, and antinuclei~\cite{DeLaTorreLuque:2024htu,Winkler:2022zdu,Cholis:2020twh} in the cosmic-ray spectrum.

Gamma-ray searches targeting different regions of the sky offer various advantages and challenges. 
The region surrounding the Galactic Center is expected to yield the brightest DM annihilation signal due to its high DM density and relative proximity to Earth. This region, however, is also characterized by large backgrounds associated with resolved and unresolved gamma-ray point sources and from cosmic rays interacting with gas and radiation in the interstellar medium~\cite{Gondolo:1999ef}. By contrast, dwarf galaxies provide comparatively low-background environments for observing DM annihilation, but this advantage is largely offset by the fact that even the most promising dwarf galaxies are predicted to yield a dramatically lower flux of DM annihilation products. For DM with a velocity-independent annihilation cross section, the gamma-ray flux predicted from dwarf galaxies is typically $\sim 10^4$ times fainter than the corresponding signal from the Galactic Center. In light of these relative advantages and disadvantages, it is likely that a DM annihilation signal would first be detected from the direction of the Galactic Center but might then require a subsequent detection from dwarf galaxies in order to robustly confirm its DM origin.

This is the situation that we currently face regarding the signal known as the Galactic Center Gamma-Ray Excess (GCE). The GCE is bright and highly statistically significant~\cite{Goodenough:2009gk,Hooper:2010mq,Hooper:2011ti,Abazajian:2012pn,Hooper:2013rwa,Gordon:2013vta,Daylan:2014rsa,Calore:2014xka,Fermi-LAT:2015sau,Fermi-LAT:2017opo}, is robust to variations in the astrophysical diffuse emission model, and has not been attributed to any known astrophysical sources or mechanisms~\cite{Cholis:2021rpp,DiMauro:2021raz}. Intriguingly, the observed spectrum and angular distribution of this signal are both consistent with arising from the annihilation of $\sim 50 \ \GeV$ DM particles. Furthermore, the overall intensity of the GCE requires these DM particles to have an annihilation cross section of approximately $\sigv \sim 10^{-26} \, \cmps$, consistent with the thermal relic prediction. The leading astrophysical explanation for the GCE is that it could potentially arise from a very large population of faint and centrally-located millisecond pulsars~\cite{Hooper:2010mq,Abazajian:2010zy,Hooper:2011ti,Abazajian:2012pn,Gordon:2013vta,Cholis:2014lta,Yuan:2014rca,Petrovic:2014xra,Brandt:2015ula,Malyshev:2024obk,Kuvatova:2024bdn}. 

It has been argued that gamma-ray observations of dwarf galaxies will be required to ultimately settle the question of whether the GCE is generated by annihilating DM. The absence of such signals has already been used to constrain values of the DM annihilation cross section motivated by thermal freeze out for DM masses in the range of tens of GeV~\cite{DiMauro:2022hue,McDaniel:2023bju}. As the {Fermi Gamma-Ray Space Telescope} continues to collect data, and new dwarf galaxies are discovered by the Rubin Observatory and other surveys, the sensitivity of these probes is expected to improve~\cite{He:2013jza,LSSTDarkMatterGroup:2019mwo}. 
In the more distant future, telescopes such as the proposed {Advanced Particle-astrophysics Telescope} (APT)~\cite{Alnussirat:2021tlo,APT:2021lhj}
will be sensitive to dwarf gamma-ray signals  from a $\sim 50 \ \GeV$ DM candidate with an annihilation cross section as small as $\sigv \sim 10^{-27}  \, \cmps$, which could confirm or falsify the DM interpretation of the GCE with very high statistical significance, at least assuming the usual relationship between dwarf and Galactic Center annihilation signals~\cite{Xu:2023zyz}. 

In this sense, dwarf galaxy observations can test the DM interpretation of the GCE under the assumption that the effective annihilation cross section is independent of the environment. 
However, the kinetic energy of a DM particle in the Inner Galaxy is approximately three orders of magnitude greater than in a dwarf galaxy. Therefore, if the effective DM annihilation rate is dependent on the velocity distribution, the relationship between the gamma-ray fluxes predicted from the Galactic Center and from dwarf galaxies will be substantially altered.

\begin{figure*}[t!]
\centering
\includegraphics[width=0.95\linewidth]{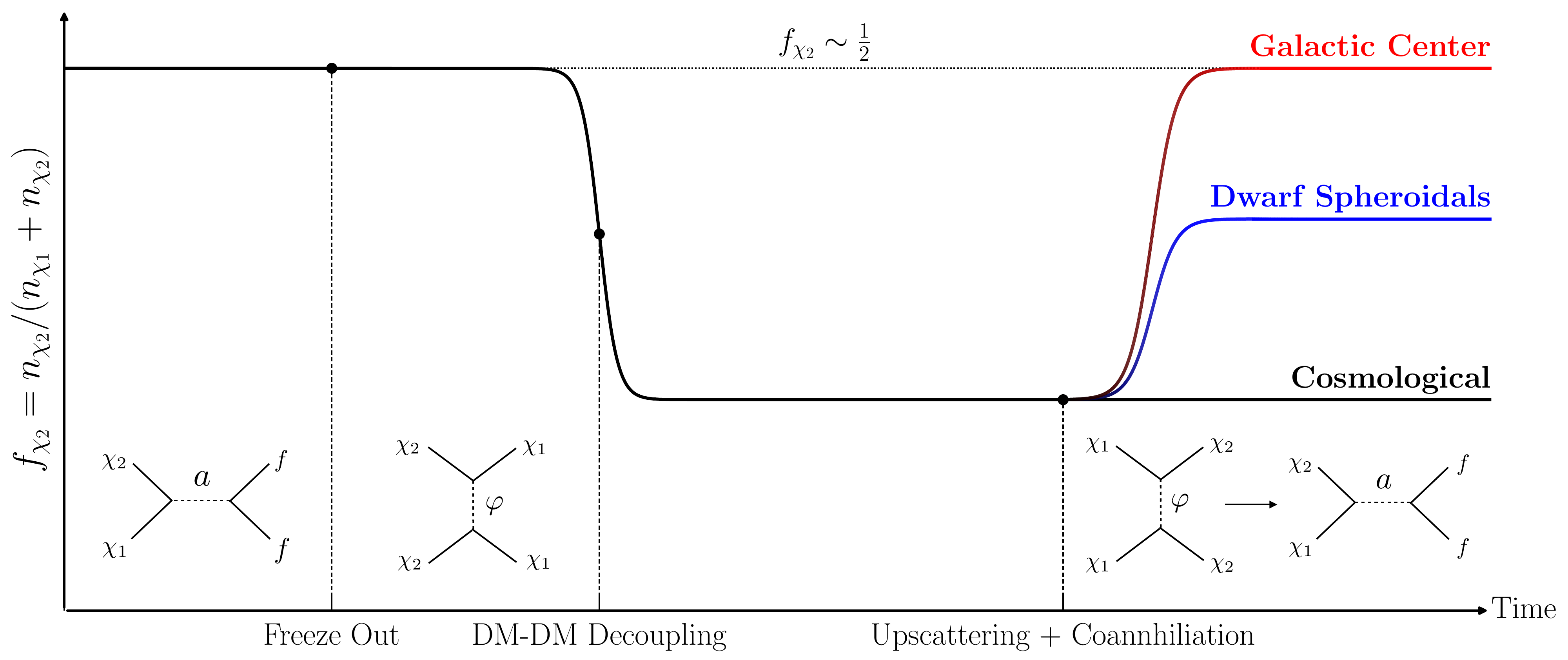}
\caption{An illustration of the key events which take place over cosmological history in the scenario presented in this study, in terms of the evolution of the fractional abundance, $f_{\x_2}$, of the heavier dark matter state, $\x_2$. After the temperature of the dark matter has become smaller than the mass splitting, $T_\x \lesssim 
\delta_\x$, downscattering depletes the primordial $\x_2$ abundance. At late times, upscattering occurs efficiently in the Galactic Halo (but not in dwarf galaxies) creating a new $\x_2$ population which generates gamma-rays through dark matter coannihilations.}
\label{fig:cartoon}
\end{figure*}

In this paper, we explore this possibility and identify a class of models in which the DM annihilation rate in dwarf galaxies is parametrically suppressed relative to the naive expectation based on the annihilation rate in the Galactic Center. 
In this scenario, gamma-ray signals are generated through the coannihilations of a ground state DM particle, $\x_1$, and a slightly heavier excited state, $\x_2$, whose primordial abundance is comparatively suppressed.
As shown in \Fig{cartoon}, DM particles in the Galactic Halo can be efficiently upscattered into the excited state, $\x_1 \x_1 \rightarrow \x_2 \x_2$, while this process remains suppressed in the lower-velocity enviroments of dwarf galaxies. 
For a range of model parameters, these dynamics predict a detectable DM coannihilation signal from the Galactic Center, at a level comparable to that expected from DM in the form of a standard thermal relic with a velocity-independent annihilation cross section, while suppressing any corresponding signal from dwarf galaxies or from the epoch of recombination.
From this perspective, if next-generation telescopes do not observe gamma-ray signals from dwarf galaxies, it would not strictly rule out a DM interpretation of the GCE.
We note that analogous ideas involving the late-time annihilation or coannihilation signals of a two-state DM system have been investigated previously in Refs.~\cite{Finkbeiner:2007kk,Finkbeiner:2014sja,Tulin:2012re,Hardy:2014dea,Berlin:2023qco}. Compared to those works, the scenario presented here is most similar to Ref.~\cite{Berlin:2023qco}, which focused on the future sensitivity to sub-GeV DM signals that arise only from low-redshift galaxy halos. 

This paper is organized as follows. In \Sec{model}, we present the main ingredients necessary to realize our scenario. In \Sec{cosmo}, we describe the cosmological evolution of the $\x_1$-$\x_2$ system and, in \Sec{upscatter}, we calculate the fraction of the DM particles that will be in the excited state in various astrophysical environments. In \Sec{IDD}, we calculate the corresponding gamma-ray fluxes from different indirect detection targets and, in \Sec{conclusion}, we offer some concluding remarks.

\section{The Dark Matter Scenario}
\label{sec:model}

We begin by describing the features that are required of a model that can generate a detectable gamma-ray signal from DM annihilation in the Galactic Halo while also suppressing the corresponding gamma-ray flux from dwarf galaxies. Following a conceptual overview, we provide an explicit construction of such a model in the context of a two-Higgs-doublet scenario. We emphasize, however, that the relevant phenomenology could potentially be realized within a broader class of DM models.

\subsection{Conceptual Overview}

As noted above, we will focus on a model in which the DM consists of two distinct states, $\x_1$ and $\x_2$, with a small mass splitting (commonly referred to as inelastic DM~\cite{Tucker-Smith:2001myb}). A gamma-ray signal is generated in this model only through the coannihilations of these two states, requiring both to be present. As schematically depicted in Fig. \ref{fig:cartoon}, the following sequence of events play out over cosmological history in this model:
\begin{itemize}
\item
{\bf Freeze out via heavy-light coannihilation} \\
The process of $\x_1$-$\x_2$ coannihilation to SM particles establishes the DM abundance.

\item
{\bf Heavy state is cosmologically depleted} \\
The process of $\x_2 \x_2 \rightarrow \x_1 \x_1$ downscattering depletes the $\x_2$ abundance, leaving the DM to consist overwhelmingly of particles in the ground state, $\x_1$. 

\item
{\bf Dark matter in the Galactic Halo is excited into the heavy state} \\ 
At late times, after structure forms, an appreciable $\x_2$ population is regenerated in virialized objects through $\x_1 \x_1 \rightarrow \x_2 \x_2$ upscattering. Because $\x_2$ is heavier, these reactions are sensitive to the velocity distribution within a given astrophysical system.
Thus, the resulting $\x_2$ population can be sizable in the Galactic Halo, but parametrically suppressed in the lower-velocity environments of dwarf galaxies.

\item
{\bf  Heavy-light coannihilation in the Galactic Halo } \\
DM in the Galactic Halo produces gamma-rays through the process of $\x_1$-$\x_2$ coannihilation to SM particles. Without a significant $\x_2$ population present in dwarf galaxies, the gamma-ray flux from those systems is highly suppressed.
\end{itemize}

In this discussion, it is useful to keep in mind the characteristic DM velocity dispersion in dwarf spheroidal galaxies (dSph), in the Milky Way (MW) halo, and at the time of thermal freeze out (FO):
\be
v_\text{dSph}^2 \sim 10^{-9} ~~,~~ 
v_{\rm MW}^2 \sim 10^{-6}~~,~~
v_{\rm FO}^2 \sim 10^{-1}~.
\ee
Dark matter coannihilation can only occur in regions where a sizable population of $\x_2$ particles has been generated through $\x_1 \x_1 \to \x_2 \x_2$ upscattering. These reactions are only possible if the kinetic energy of the DM exceeds the mass splitting between these two states:
\be
\ml v^2 \gtrsim \mh - \ml \equiv \delta_\x
~,
\ee
where $v$ is the relative velocity of the initial state particles, $\x_1$, and we have defined the mass splitting, $\delta_\x$. We therefore focus on splittings of size
\be
\label{eq:Delta1}
v_\text{dSph}^2  \ll \frac{\delta_\x}{\ml} \ll v_{\rm MW}^2 
~,
\ee
such that DM particles can be efficiently upscattered in the halo of the Milky Way, while this process is kinematically suppressed in dwarf galaxies.

\subsection{Concrete Realization}

Here we present a scenario whose minimal phenomenological ingredients consist of a nearly-degenerate pair of Majorana fermions, $\x_1$ and $\x_2$, split in mass by $\delta_\x = \mh - \ml \ll m_{\x_{1,2}}$, as well as two mediators, corresponding to a pseudoscalar, $a$, and a scalar, $\varphi$, with masses $m_{\varphi} \ll m_{\x_{1,2}} < m_a$. The first mediator, $a$, is responsible for the process of DM coannihilation. The second, $\varphi$, is necessary to facilitate large DM upscattering rates at late times. Unlike $a$,  $\varphi$ does not couple directly to the SM. The relevant interactions can be parametrized with the Lagrangian,
\be
\label{eq:toy-lag}
\Lag_{\rm int} =  y_\x^\prime \, \varphi \, \bar \x_1 \x_2 + y_\x \, a \, \bar\x_1 i\gamma^5 \x_2 + 
y_f \, a \, \bar f i\gamma^5 f
~, 
\ee
where $y_\x^\prime$, $y_\x$, and $y_f$ are dimensionless Yukawa couplings and $f$ is a SM fermion. Since $m_\varphi \ll m_a$, $\x_1 \x_1 \leftrightarrow \x_2 \x_2$ scattering reactions predominantly proceed through $\varphi$-exchange, which is governed by the first term in \Eq{toy-lag}. As we show in \Sec{cosmo}, this process is efficient in the early and late universe for $y_\x^\p \gg 10^{-2}$. The last two terms of \Eq{toy-lag} instead dictate $\x_1 \x_2 \leftrightarrow f \bar{f}$ coannihilation reactions, which proceed through the exchange of an off-shell $a$.\footnote{The general features we describe here do not strictly require two different particles to mediate the annihilation and scattering processes. For example, the model in Ref.~\cite{Berlin:2023qco} exhibits similar scattering and annihilation dynamics for MeV-scale DM utilizing only one mediator particle. However, if the same model were to be realized with weak-scale DM masses, the cross sections needed for non-trivial Galactic upscattering rates would require non-perturbative couplings. Thus, here we invoke an additional lighter mediator to enable viable upscattering in the Galactic Halo.} For this latter process to establish the DM abundance of weak-scale DM, $y_\x \sim 1 \gg y_\x^\p$ is needed (see \Sec{cosmo}). For this hierarchy of couplings, there is a predictive relationship between the thermal relic abundance and the magnitude of the indirect direction signal.

In the remainder of this section, we outline a complete gauge-invariant model that realizes these ingredients, with the off-diagonal $\x_{1,2}$ interactions of \Eq{toy-lag} arising naturally from discrete symmetries. We emphasize, however, that any model with the phenomenological features outlined above would suffice to realize the dynamics of our scenario. 

\subsubsection{Mass Spectrum}

To generate masses for the $\x_1$ and $\x_2$, we introduce a pair of singlet Weyl spinors, $\x$ and $\x^c$, and impose upon the theory a charge-conjugation symmetry, $\mathcal{C}_\x$, under which $\x \leftrightarrow \x^c$, as well as an approximate $\mathbb{Z}_2 \times \mathbb{Z}_2^c$ symmetry under which $\x$ and $\x^c$ separately transform as $\x \to - \x$ and $\x^c \to - \x^c$. The mass terms for these spinors are described by the Lagrangian,
\be
\label{eq:LagMass}
- \Lag \supset \frac{M}{2} \left(  \x \x + \x^{c}\x^{c} \right) + m \,  \x  \x^c + \text{h.c.}
~,
\ee
where $M$ and $m$ are Majorana and Dirac masses, respectively. Since the latter softly breaks $\mathbb{Z}_2 \times \mathbb{Z}_2^c$ to its diagonal subgroup, it is technically natural to take $m \ll M$. The Majorana mass eigenstates of \Eq{LagMass} are
\be
\x_{1,2} = \frac{1}{\sqrt{2}} \, \big( \x \mp \x^c \big)
~,
\ee
which are nearly degenerate with masses $m_{\x_{1,2}} = M \mp m$ and a mass splitting of $\delta_\x = 2 m$. In the remainder of this work, we approximate $m_\x \equiv \ml \approx \mh$ and expand in the small parameter $\delta_\x/m_\x \ll 1$.

\subsubsection{Coannihilation Interaction}

To obtain the interactions of the pseudoscalar mediator, we begin by introducing a gauge singlet pseudoscalar, $a_0$, that is odd under the $\mathcal{C}_\x$ symmetry introduced above. Since $a_0$ is also odd under parity, the only allowed coupling between $a_0$ and the DM sector is
\be
\label{eq:PScoupling}
\Lag \supset 
- i  \, \frac{y_\x}{2} \, a_0 \, \big( \x \x - \x^c \x^c \big)+ \text{h.c.} 
\ee
When written in terms of the mass eigenstates, $\x_{1,2}$, this interaction is purely off-diagonal and takes the form of the second term in \Eq{toy-lag}, after writing Eq. \ref{eq:PScoupling} in four-component notation.\footnote{Since the mediator, $a_0$, couples to the gauge eigenstates of this theory, oscillations between $\x_{1}$ and $\x_2$ can occur~\cite{Buckley:2011ye}. Like with neutrino oscillations, however, we can safely treat $\x_1$ and $\x_2$ as distinguishable mass eigenstates if the oscillation rate is sufficiently fast. For a non-relativistic two-level system, the oscillation rate is given by $\delta_\x$. Demanding that this is greater than the rate of Hubble expansion, we find that the temperature at which $\x$ and $\x^c$ decohere is $T \sim \sqrt{M_{\rm Pl} \, \delta_\x}$. Throughout the parameter space of interest here, this occurs well before the time of DM freeze out, allowing us to safely ignore any such quantum effects.} 

The coupling of the mediator, $a_0$, to the SM closely follows the setup described in Ref.~\cite{Ipek:2014gua}, which features  a Type-II two-Higgs-doublet model (for a review, see Ref.~\cite{Branco:2011iw}). In these models, the two Higgs doublets are parametrized as
\be
\Phi_{u,d} = \mat{
H_{u,d}^+ \\
\frac{1}{\sqrt{2}} \, \big( v_{u,d} + h_{u,d} + ia_{u,d} \big)
}
~,
\ee
where $\Phi_u$ couples to the up-type quarks, and $\Phi_d$ couples to the down-type quarks and charged leptons. Two different linear combinations of $H_{u,d}^+$ pair up to become a charged Higgs boson, $H^\pm$, and a Goldstone boson that is eaten by the $W^\pm$. Two different linear combinations of $h_{u,d}$ pair up to become a light and heavy CP-even Higgs scalar, $h$ and $H$. Finally, two different linear combinations of $a_{u,d}$ pair up to become a CP-odd pseudoscalar, $A_0$, and a Goldstone boson that is eaten by the $Z$. The vacuum expectation values of the two Higgs doublets are related to that of the SM according to $v^2_h = v_u^2 + v_d^2 \approx (246 \ \GeV)^2$, and we define $\tan{\beta} \equiv v_u / v_d$. Here, we will work in the alignment limit, in which $h$ possesses the same couplings as the SM Higgs boson.

Under a CP transformation, $\Phi_{u,d} \to \Phi_{u,d}^*$ and $a_0 \to - a_0$. The relevant CP-invariant mass terms of the Higgs sector are thus given by
\be
\Lag \supset - \frac{1}{2} \, m_{A_0}^2 \, A_0^2 - \frac{1}{2} \, m_{a_0}^2 \, a_0^2 - (i B \, a_0 \, \Phi_d^\dagger \Phi_u + \text{h.c.})
~.
\ee
In the expression above, $m_{A_0}$ and $m_{a_0}$ are the masses of $A_0$ and $a_0$, respectively, in the absence of mixing, and the term involving the dimensionful coupling, $B$, describes the leading CP-conserving gauge-invariant interaction between $a_0$ and $\Phi_{u,d}$.\footnote{Note that a term of the form $a_0^2 \, |\Phi_{u,d}|^2$ is also allowed, but has no phenomenological consequences for the signals discussed in this work.} Note that since $B \neq 0$ softly breaks the charge-conjugation symmetry $\mathcal{C}_\x$ presented above, it is natural to expect $B \lesssim 1 \ \TeV$. After electroweak symmetry breaking, and writing $a_{u,d}$ in terms of the physical 2HDM pseudoscalar, $a_u \to \cos{\beta} \, A_0$ and $a_d \to - \sin{\beta} \, A_0$~\cite{Branco:2011iw}, the above mass terms become 
\be
\Lag \supset - \frac{1}{2} \, m_{A_0}^2 \, A_0^2 - \frac{1}{2} \, m_{a_0}^2 \, a_0^2 + B \, v_h \, A_0 \, a_0
~,
\ee
corresponding to a mass-mixing between $A_0$ and $a_0$. To leading order in $m_{A_0}^2 \gg m_{a_0}^2, B v_h$, the mass-diagonal basis is obtained by $A_0 \to A + \theta a$ and $a_0 \to a - \theta A$, where $a$ is the lighter $a_0$-like pseudoscalar, $A$ is the heavier $A_0$-like pseudoscalar, and the mixing angle is approximately
\be
\theta \approx B \, v_h / m_{A_0}^2
~.
\ee
We are thus left with a light pseudoscalar, $a$, which couples to the DM as described in \Eq{PScoupling}, and to the SM as the $A_0$ does, but with couplings that are scaled down by a factor of $\theta$: 
\be
\Lag \supset y_f\, a \, \bar{f} i \g^5 f
~,
\ee
which is identical to the last term of \Eq{toy-lag}, and where the coupling is given by
\be
y_f \approx \theta \, \frac{m_f}{v_h}
\times 
\begin{cases}
\cot{\beta} & (\text{up-type quarks})
\\
\tan{\beta} & (\text{down-type quarks,  leptons})
~,
\end{cases}
\ee
and $m_f$ is the SM fermion mass. The mass of this mediator is given by $m_a^2 \approx m_{a_0}^2 - \theta^2 \, m_{A_0}^2$. The perturbative limit thus corresponds to $\theta \lesssim m_{a_0} / m_{A_0}$. Note that due to Bose symmetry, there are no trilinear interactions between $A_0$ and a pair of vector bosons, and furthermore, in the alignment limit, the $A_0 h Z$ interaction vanishes~\cite{Gunion:1989we}. 

In this work, we will focus on the $\tan{\beta} \gg 1$ limit, in which case the pseudoscalar, $a$, dominantly couples to bottom quarks and tau leptons. Perturbativity of the bottom quark coupling, $m_b / v_d \lesssim 1$, imposes an upper bound of $\tan{\beta} \lesssim v_h / m_b \sim 60$. Stronger bounds arise from direct searches for $A_0 \to \tau^+ \tau^-$ at the LHC, which constrain $\tan{\beta} \lesssim 10 - 50$ for $m_{A_0} \sim (1-2) \ \TeV$~\cite{ATLAS:2020zms}. We also note that modifications to the observed leptonic decay of $B$-mesons constrain $\theta \lesssim 0.1$ for $m_a \sim 10^2 \ \GeV$~\cite{Ipek:2014gua}. For the remainder of this work, we will adopt $m_\x=50 \ \GeV$, $m_a = 150 \ \GeV$, $\theta = 0.1$, and $\tan{\beta} = 10$, as listed in Table~\ref{tab:parm}.

\begin{table*}[t]
\ra{1.3}
\centering
\tabcolsep=0.2cm
\begin{tabularx}{0.5\textwidth}{@{\extracolsep{3pt}}cccccc@{}}
\hlinewd{1pt}
\multicolumn{1}{c}{Parameter}  & \multicolumn{1}{c}{$m_\chi$} & \multicolumn{1}{c}{$m_a$} & \multicolumn{1}{c}{$\theta$} & \multicolumn{1}{c}{$\tan\beta$} & \multicolumn{1}{c}{$m_\phi / \delta_\chi$} \\ 
\hlinewd{0.5pt}
Value & $50 \ \GeV$ & $150 \ \GeV$      &  $0.1$ & $10$ & 3  \\ 
\hlinewd{1pt}
\end{tabularx}
\caption{The parameter values which we have adopted throughout this analysis. For this combination of parameters, the effective coupling of the mediator to $b$-quarks is given by $y_f \approx \theta m_b \tan \beta/v_h \approx 0.017$.}
\label{tab:parm}
\end{table*}

\subsubsection{Scattering Interaction}
As mentioned above, efficiently upscattering DM in the Galactic Halo will require us to introduce a second mediator, $\varphi$, which is much lighter than the pseudoscalar, $a$. We parameterize the interactions of this light scalar as follows:
\be
\label{eq:DS}
\Lag \supset y_\x^\p \, \varphi \,  \bar{\x}_1 \, \x_2 + \cdots
~.
\ee
Note that since $\varphi$ does not couple to the SM, it would not necessarily be problematic for terms of the form $\varphi \,  \bar{\x}_1 \x_1$ or $\varphi \,  \bar{\x}_2 \x_2$ to be present. We do, however, take these terms to be subdominant.\footnote{We note that terms of the form $\varphi \,  \bar{\x}_1 \x_1$ and $\varphi \,  \bar{\x}_2 \x_2$ would be forbidden if $\varphi$ is taken to be odd under $\mathcal{C}_\x$. In this case, the interaction in Eq.~\ref{eq:DS} arises from the parity-even analogue of \Eq{PScoupling}.} For the remainder of this work, we will adopt $m_{\varphi} = 3 \delta_\x$, as listed in Table~\ref{tab:parm}. As discussed below, this choice is motivated by the fact that, for $m_\varphi > \delta_\x$, the decay process $\x_2 \to \x_1 \varphi$ is kinematically forbidden, while for $m_\varphi \lesssim \delta_\x / v$, the   upscattering process, $\x_1 \x_1 \to \x_2 \x_2$, is enhanced by the exchange of the light $\varphi$, where $v$ is the $\x_1$ velocity. Since $v \ll 1$, our results remain largely unchanged for much larger values of $m_\varphi$.

\section{Cosmological History}
\label{sec:cosmo}

In this section, we consider the thermal freeze out of DM in this scenario, and discuss the cosmological constraints that apply to this class of models. 

\subsection{Dark Matter Freeze Out}

In the small mass splitting limit, $\delta_\x \ll m_\x$, the non-relativistic cross section for the coannihilation process, $\x_1 \x_2 \to a^* \to f \bar{f}$, is given by 
\be
\sigma v_{\x_1 \x_2 \to f \bar{f}} \approx \frac{n_f \, y_\x^2 \, y_f^2}{2 \pi} \, \frac{m_\x^2\sqrt{ 1 - m_f^2 / m_\x^2} }{\big( 4 m_\x^2 - m_a^2 \big)^2}
~,
\ee
where $n_f =1 \,(3)$ for final state leptons (quarks). For $\tan{\beta} \gg 1$, these coannihilations proceed mostly to bottom quarks, with $y_f \approx \theta \, m_b \tan \beta /v_h$. In calculating the thermal relic abundance of DM in this model, we employ the standard semi-analytic results as described in Ref.~\cite{Berlin:2016gtr}, and include all allowed annihilation channels to SM fermions.  We find that in the $m_\x \sim (10 - 100) \ \GeV$ mass range, the thermal relic abundance equals the measured DM density if $\sigma v_{\x_1 \x_2 \to b \bar{b}} \approx 4.4 \times 10^{-26} \, \cmps$~\cite{Steigman:2012nb}, corresponding to
\be
y_\x \approx 1.3 \times \bigg( \frac{0.1}{\theta} \bigg) \, \bigg( \frac{10}{\tan{\beta}} \bigg) \, \bigg( \frac{m_\x}{50 \ \GeV} \bigg) \, \bigg( \frac{m_a / m_\x}{3} \bigg)^2
~.
\ee
Throughout this work, we fix $y_\x$ such that the abundance of $\x_1$ and $\x_2$ agrees with the measured DM density.

In this model, $\x_1 \x_1$ $(\x_2 \x_2)$ can also annihilate through the $t$-channel exchange of a $\x_2$ ($\x_1$) to produce a pair of light scalar mediators, with a cross section given by
\be
\label{eq:X2PhiFO}
\sigma v_{\x_1 \x_1 \to \varphi \varphi} \approx \sigma v_{\x_2 \x_2 \to \varphi \varphi} \approx \frac{3 y^{\p\,4}_\x v^2}{128 \pi m_\x^2}
~, 
\ee
where $v$ in this expression is the relative velocity of the two annihilating particles. Throughout this study, we restrict the value of the DM-$\varphi$ coupling, $y_\x^\p$, such that the processes $\x_1 \x_1 \to \varphi \varphi$ and $\x_2 \x_2 \to \varphi \varphi$ contribute less than 10\% of the total DM annihilation rate at the temperature of thermal freeze out. This requirement is satisfied for 
\be
\label{eq:ypmax}
y^{\p}_\x \le y_\x^{\p\,{\rm max}} \approx 0.14 \times \bigg(\frac{m_\x}{50 \ \GeV}\bigg)^{1/2}
~,
\ee
which ensures that freeze out is set by  coannihilation to SM fermions, $\x_1 \x_2 \to f \bar{f}$, and thereby preserves the predictive relationship between the relic abundance and the magnitude of the Galactic indirect direction signal.

Finally, we mention the implications for direct detection in this scenario, which are studied comprehensively in Ref.~\cite{Ipek:2014gua}. In this model, the leading order process for elastic DM-nucleus scattering involves a loop of $a$ particles, which generates a scalar interaction between $\x_1$ and the SM-like Higgs boson. The resulting spin-independent, elastic scattering cross section with nucleons, $n$, satisfies
\be
\sigma_{\x n} \sim 10^{-48} \ \cm^2 \times \bigg(\frac{y_\x \, \theta \,  m_A}{m_a}\bigg)^4 \, \bigg( \frac{m_\x}{50 \ \GeV} \bigg)^2
~,
\ee
which, in our parameter space of interest, $y_\x \sim \theta \,  m_A / m_a \sim 1$, is consistent with current direct detection constraints~\cite{LZ:2024zvo}.

\subsection{Cosmological Constraints}

We have taken the secluded scalar mediator to be light in this scenario, $m_{\varphi} \lesssim \MeV$, such that this particle acts as radiation during the eras of Big Bang nucleosynthesis and possibly recombination. Note that, at tree level, $\varphi$ couples only to the DM, so its cosmological population decouples from the SM bath when $\x \x \to \varphi \varphi$ annihilation freezes out at a temperature of $T_{\rm dec} \gtrsim m_\x/20$. For weak scale DM, this occurs well before nucleosynthesis or recombination, so the temperature of the $\varphi$ population is reduced relative to that of the SM bath by a factor of $[g_{* S}(T_{\rm dec})/g_{* S}(T_{\rm CMB})]^{1/3}$ due to the SM entropy transfers from the QCD phase transition~\cite{Vogel:2013raa,Adshead:2022ovo}, where $g_{* S}$ is the effective number of relativistic degrees-of-freedom in entropy, and $g_{* S}(T_{\rm CMB})\approx 3.91$. These transfers decrease the $\varphi$ contribution to the radiation density such that the effective number of neutrino species is shifted by
\be
\Delta \Neff \approx \frac{4}{7} \, \left[ \frac{11}{4}\, \frac{g_{*S}(T_{\rm CMB}) }{g_{*S}(T_{\rm dec})} \right]^{4/3}
\approx 0.04  \times \bigg[\frac{80}{g_{*S}(T_{\rm dec})}\bigg]^{4/3}
~,
\ee
which is consistent with current constraints~\cite{Calabrese:2025mza,Planck:2018vyg}.

\subsection{Downscattering}
\label{sec:downscattering0}

In the early universe, once the DM temperature drops below the mass splitting, $T_\x \lesssim  \delta_\x$, the $\x_2$ population is depleted through the downscattering processes, $\x_2 \x_2 \to \x_1 \x_1$ and $\x_2 f \rightarrow \x_1 f$, which occur through $\varphi$ and $a$ exchange, respectively. Since $m_\varphi \ll m_a$, the former process dominates over the latter. The thermally-averaged rate for this process is approximately given by (see Appendix~\ref{sec:downscattering})
\be
\label{eq:downscatter}
\Gamma_{\x_2 \x_2 \to \x_1 \x_1} \approx 
\, \frac{g_{*S} (T_\text{SM})}{g_{* S}(T_\text{eq})} \, \frac{T_\text{eq} \, T_\text{SM}^3}{2 m_\x} \, f_{\x_2}
\, \langle \sigma v \rangle_{\x_2 \x_2 \to \x_1 \x_1} 
~,
\ee
where $T_\text{SM}$ is the temperature of the SM bath, $T_\text{eq} \approx 0.8 \ \eV$ is the SM temperature at matter-radiation equality, and $f_{\x_2}$ is the relative density of $\x_2$ particles,
\be
\label{eq:fx2}
f_{\x_2} \equiv \frac{n_{\x_2}}{n_{\x_1}+n_{\x_2}}= \frac{e^{- \delta_\x / T_\x}}{1 + e^{-\delta_\x / T_\x}}
~,
\ee
where $T_\x$ is the DM temperature. In the $m_\varphi \ll m_\x$ limit, the thermally-averaged cross section (accounting for the Sommerfeld enhancement from the long-ranged $\varphi$-potential~\cite{Cirelli:2016rnw}) is
\be
\label{eq:sigmavdown}
\langle \sigma v \rangle_{\x_2 \x_2 \to \x_1 \x_1}  \approx \frac{y_\x^{\p \, 6}}{2^{5/2} \pi^2 \, \delta_\x^2}
~.
\ee
The scaling with $\delta_\x^{-2}$ implies that this process is significantly enhanced by the small mass splittings considered in this work. 
As discussed in Appendix~\ref{sec:downscattering}, this enhancement arises from forward scattering events that exchange a small amount of momentum, and is regulated by the $\varphi$ mass, saturating if the mass exceeds the minimum momentum transfer, $m_{\varphi} \gtrsim \delta_\x / v$.

This downscattering process decouples once the rate described in \Eq{downscatter} falls below the rate of Hubble expansion, $H$. As a result, the $\x_2$ cosmological number density is efficiently depleted if this occurs after the DM temperature drops below the mass splitting, $T_\x \lesssim \delta_\x$, corresponding to 
\be
\label{eq:yxpmin}
y_\x^\p \gtrsim 10^{-3} \times  \bigg( \frac{m_\x}{50 \ \GeV}\bigg)^{1/3} \, \bigg( \frac{T_\x / T_\text{SM}}{10^{-3}} \bigg)^{1/6} \, \bigg( \frac{\delta_\x / m_\x}{10^{-8}} \bigg)^{1/6}
~.
\ee
Note that $T_\x / T_{\rm SM} \ll 1$ due to the fact that the DM temperature evolves as $T_\x \propto T_\text{SM}^2$ after the DM undergoes kinetic decoupling, which typically occurs well before the process $\x_2 \x_2 \leftrightarrow \x_1 \x_1$ chemically decouples (see Appendix~\ref{sec:tempevol}). Comparing the above expression to \Eq{ypmax}, we see that there is a wide range of $\varphi$ couplings, $y_\x^\p \sim 10^{-3} - 10^{-1}$, for which $\varphi$-exchange can significantly deplete the relative primordial DM fraction of $\x_2$ particles without significantly altering the total DM abundance through $\x_{1} \x_{1} \to \varphi \varphi$ or $\x_{2} \x_{2} \to \varphi \varphi$. 

\begin{figure}[t!]
\centering
\includegraphics[width=0.6\columnwidth]{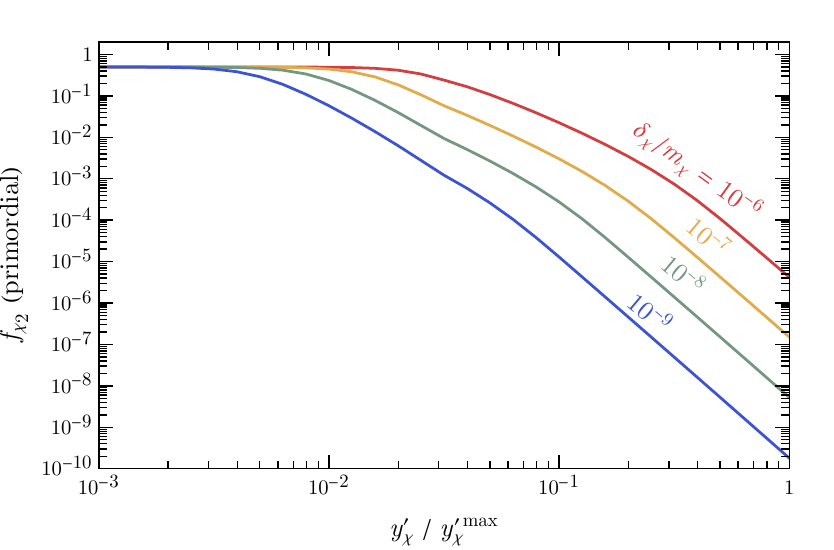}
\caption{The primordial $\x_2$ fraction, $f_{\x_2} \equiv n_{\x_2}/(n_{
\x_1}+n_{\x_2})$, as a function of $y_\x^\p$, normalized by $y_\x^{\p \, \text{max}}$ (see \Eq{ypmax}), for the model parameters given in Table~\ref{tab:parm}, and for various choices of the dark matter mass splitting, $\delta_\x$.}
\label{fig:f2yxp}
\end{figure}

To determine the relative fraction of the DM that is composed of $\x_2$ particles, we use the instantaneous freeze-out approximation. In this case, the relative number density of $\x_2$ particles is given by \Eq{fx2}, evaluated with the DM temperature set to the value at which $\Gamma_{\x_2 \x_2 \to \x_1 \x_1} = H$. In \Fig{f2yxp}, we show this primordial $\x_2$ fraction as a function of $y_\x^\p$ for representative model parameters and various choices of the DM mass splitting. In agreement with the analytic form of \Eq{yxpmin}, this shows that for $y_\x^\p \gtrsim 10^{-2} \times y_\x^{\p \, \text{max}} \sim 10^{-3}$, the primordial abundance of $\x_2$ is efficiently depleted through downscattering. Also note that for smaller values of the mass splitting, $\delta_\x$, the $\x_2$ fraction is reduced, due to the enhancement described by \Eq{sigmavdown}.

\section{Late-Time Upscattering}
\label{sec:upscatter}

As shown above, the primordial abundance of $\x_2$ at early times is efficiently depleted through downscattering, $\x_2 \x_2 \to \x_1 \x_1$, such that at late times the DM abundance consists mostly of $\x_1$ particles. Therefore, to generate a gamma-ray signal through the coannihilation process, $\x_1 \x_2 \rightarrow f\bar{f}$, a $\x_2$ population must first be regenerated through the upscattering of DM particles into the excited state, $\x_1 \x_1 \to \x_2 \x_2$. As in the previous section, this process proceeds through the exchange of the light scalar mediator, $\varphi$, with a rate that is given by
\be
\label{eq:upscatter}
\Gamma_{\x_1 \x_1 \to \x_2 \x_2} \approx \frac{\rho_\x}{m_\x} \, e^{-2 \delta_\x / T_\x} \, \langle \sigma v \rangle_{\x_2 \x_2 \to \x_1 \x_1}
~,
\ee
where we have used detailed balance to express this in terms of the downscattering cross section given in \Eq{sigmavdown}, and $\rho_\x$ is the DM density in an astrophysical system, such as the halo of the Milky Way. Note that this form of the rate assumes a Boltzmann distribution for the DM phase space, specified by the effective temperature, $T_\x$. Demanding that the time for a DM particle to upscatter is less than the age of the universe yields the following condition:
\be
y^\p_\x \gtrsim 10^{-2} \times e^{\delta_\x / 3 T_\x} \,  \bigg(\frac{m_\x}{50 \ \GeV}\bigg)^{1/2} \, \bigg( \frac{\delta_\x / m_\x}{10^{-8}}\bigg)^{1/3} \, \bigg(\frac{10 \ \GeV / \cm^3}{\rho_\x}\bigg)^{1/6}.
\ee
Therefore, for astrophysical environments in which $T_\x \gtrsim \delta_\x$, DM upscattering can efficiently repopulate a late-time abundance of $\x_2$ particles.

We begin by treating the DM population as a single isolated system, which is a good approximation in the case of the Galactic Halo. To determine the regenerated $\x_2$ abundance, we solve the following Boltzmann equations:
\be
\label{eq:Boltzmann1}
\dot{f}_{\x_2} = \Big( f_{\x_1}^2 \, e^{- 2 \delta_\x / T_\x} - f_{\x_2}^2 \Big) \, \frac{\rho_\x}{m_\x} \, \sigv_{\x_2 \x_2 \to \x_1 \x_1}
\quad,\quad
\dot{f}_{\x_1} = - \dot{f}_{\x_2}
~,
\ee
where $f_{\x_1} = 1 - f_{\x_2}$ is the relative density of $\x_1$ particles, and we have used detailed balance to simplify the form of the first term of the first equation. The exponential factor in this expression implies that a late-time density of $\x_2$ particles is regenerated in appreciable numbers only in astrophysical structures with $T_\x \gtrsim \delta_\x$. We solve these equations numerically using the primordial value of $f_{\x_2}$ as an initial condition, as determined by the analysis described in \Sec{downscattering0}.

\begin{figure}[t]
\centering
\includegraphics[width=0.49\columnwidth]{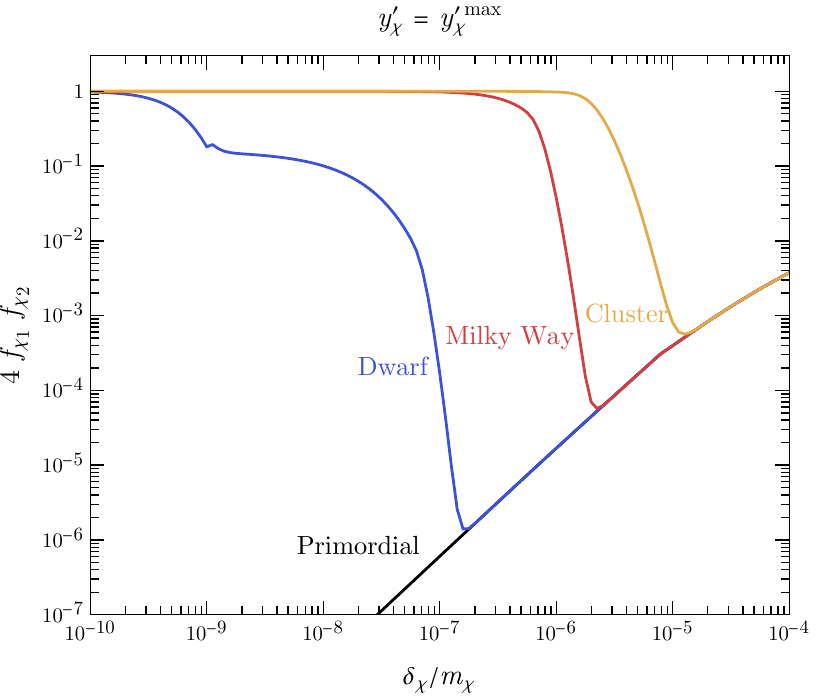}
\includegraphics[width=0.49\columnwidth]{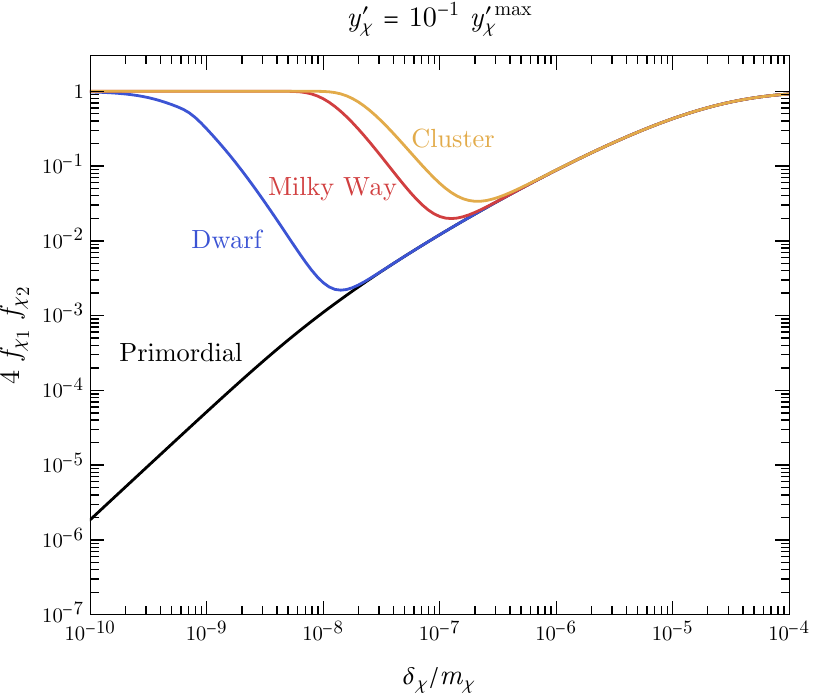}
\caption{The product of fractional abundances, $f_{\x_i} \equiv n_{\x_i}/(n_{\x_1}+n_{\x_2})$, plotted against the dark matter fractional mass splitting, $\delta_\x/m_\x$, for representative density and velocity parameters in dwarf galaxies, the Milky Way, and galaxy clusters. In particular, we adopt a dark matter mass density of  $\rho_\x = 1 \ \GeV / \cm^3$ and dark matter temperatures, $T_\x = 10^{-5} \, m_\x  / 3$, $T_\x = 10^{-6} \, m_\x  / 3$, and $T_\x = 10^{-9} \, m_\x  / 3$, corresponding to those found in galaxy clusters, in the halo of the Milky Way, and in dwarf galaxies, respectively. Furthermore, to incorporate interactions between dark matter particles in dwarf galaxies and in the halo of the Milky Way, we have assumed an ambient Galactic density of $\rho_\x^\text{MW} \approx 4 \times 10^{-3} \ \GeV / \cm^3$ near the position of the dwarf galaxy (see Appendix~\ref{sec:sub}). In both panels, we adopt the benchmark model parameters from Table~\ref{tab:parm}. In the left (right) frame, we show results for $y_\x^\p = y_\x^{\p\, \text{max}}$ ($y_\x^\p = 10^{-1} \ y_\x^{\p\, \text{max}}$), where $y_\x^{\p\,{\rm max}}$ is the value of this coupling for which the processes $\x_1 \x_1 \to \varphi \varphi$ and $\x_2 \x_2 \to \varphi \varphi$ contribute 10\% of the total annihilation rate at the time of thermal freeze out. The black curves are defined using the initial primordial values of $f_{\x_i}$ before late-time upscattering regenerated a population of $\x_2$ particles. These results confirm that, for these benchmarks, upscattering can efficiently repopulate the $\x_2$ population in galaxies and clusters, but not in dwarf galaxies.}
\label{fig:f1f2}
\end{figure}

Whereas Eq.~\ref{eq:Boltzmann1} describes the evolution of an isolated DM system, this is not necessarily a good approximation for dwarf galaxies, which are embedded within the larger halo of the Milky Way. In such satellite halos, slowly moving $\x_1$ particles can upscatter with the higher-velocity $\x_1$ population of the Galactic Halo, even if $\delta_\x/m_\x \gtrsim v_\text{dSph}^2 \sim 10^{-9}$. For an upscattered $\x_2$ particle to remain bound to its host dwarf galaxy, the imparted velocity must not exceed the dwarf escape velocity, which requires $2\delta_{\x}/(m_{\x}v_{\rm MW}) \lesssim v_{\rm dSph}$, or equivalently $\delta_{\x}/m_{\x} \lesssim 10^{-8}$. Thus, scattering between dwarf and Milky Way DM populations can significantly increase the range of mass-splittings under which $\x_2$ particles are regenerated in dwarf galaxies. We describe this aspect of our calculation in more detail in Appendix~\ref{sec:sub}. 

Once again, we solve the Boltzmann equations numerically using the primordial value of $f_{\x_2}$ as an initial condition. As a representative example, we adopt a DM density of  $\rho_\x = 1 \ \GeV / \cm^3$ and various DM temperatures, $T_\x = 10^{-5} \, m_\x  / 3$, $T_\x = 10^{-6} \, m_\x  / 3$, and $T_\x = 10^{-9} \, m_\x  / 3$, corresponding to those found in galaxy clusters, in the halo of the Milky Way, and in dwarf galaxies, respectively. Furthermore, to incorporate interactions between DM particles in dwarf galaxies and in the halo of the Milky Way, we assume an ambient Galactic density of $\rho_\x^\text{MW} \approx 4 \times 10^{-3} \ \GeV / \cm^3$ near the position of the dwarf galaxy. These results are shown in \Fig{f1f2}, for a representative choice of model parameters. Here, we plot the quantity, $4 f_{\x_1} f_{\x_2}$, in order to facilitate a straightforward comparison with the scenario of conventional Dirac DM, in which case $f_{\x_1} = f_{\x_2} = 1/2$ and $4 f_{\x_1} f_{\x_2} = 1$. These results confirm that over the large range of mass splittings given in \Eq{Delta1}, upscattering can be efficient in producing a large $\x_2$ population in the Galaxy and galaxy clusters, but not in dwarf galaxies. 

Even if DM upscattering is very efficient in the Milky Way halo,  $\x_2$ must be very long-lived in order to produce a detectable gamma-ray signal through coannihilations with $\x_1$ at late times. In the parameter space of interest, $\x_2$ decays proceed through an off-shell pseudoscalar, $a$, into a pair of photons.  As described in Appendix~\ref{sec:decays}, the Feynman diagram for the decay process, $\x_2 \to \x_1 \g \g$, involves a loop of quarks or charged leptons. For the parameter space of interest in this paper, the corresponding $\x_2$ lifetime exceeds the age of the universe
by more than twenty orders of magnitude, and is therefore unconstrained by any existing limits on DM decays~\cite{Essig:2013goa}.

Before concluding this section, we will also comment on the effect of momentum-transfer through the same upscattering process as considered above, $\x_1 \x_1 \to \x_2 \x_2$, which is bounded from considerations of self-interacting DM (SIDM) and its effect on the mass profile of various astrophysical systems~\cite{Tulin:2017ara}. For identical initial state particles, the relevant quantity of interest dictating the momentum-exchange rate is the viscosity cross section, $\sigma_V$. As discussed in Appendix~\ref{sec:downscattering}, unlike the upscattering cross section described in \Eq{sigmavdown}, the enhancement of $\sigma_V$ for small mass splittings is significantly softened by the fact that forward scattering does not meaningfully impact the mass profile of DM halos. As a result, even for $y_\x^\p = y_\x^{\p \, \text{max}}$, we have $\sigma_V / m_\x \lesssim \text{few} \times \cm^2 / \text{g}$ in galaxies and galaxy clusters, which is consistent with existing constraints~\cite{Tulin:2017ara}.

\section{Indirect Detection}
\label{sec:IDD}

\begin{figure}[t!]
\centering
\includegraphics[width=0.49\columnwidth]{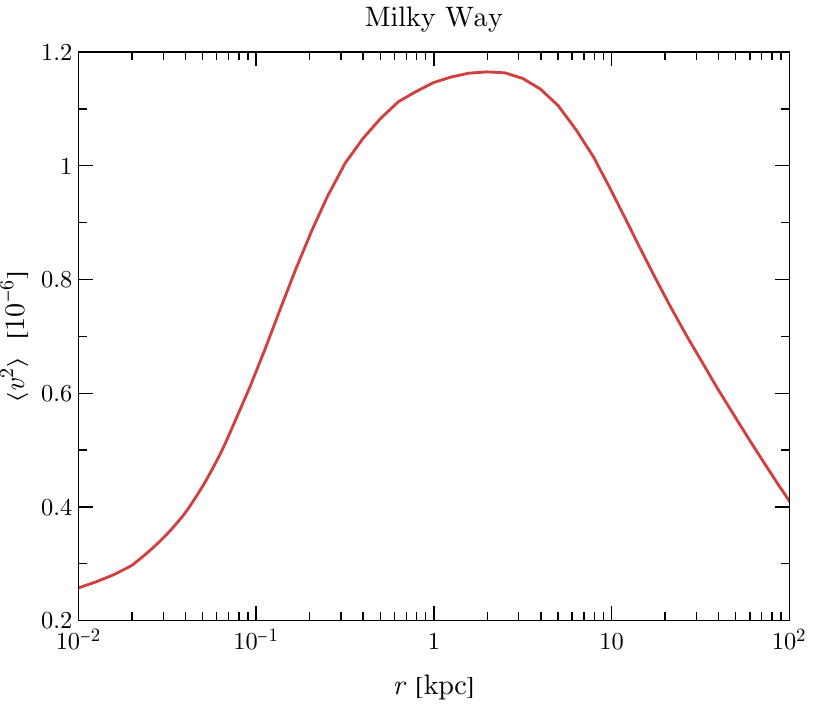}
\includegraphics[width=0.48\columnwidth]{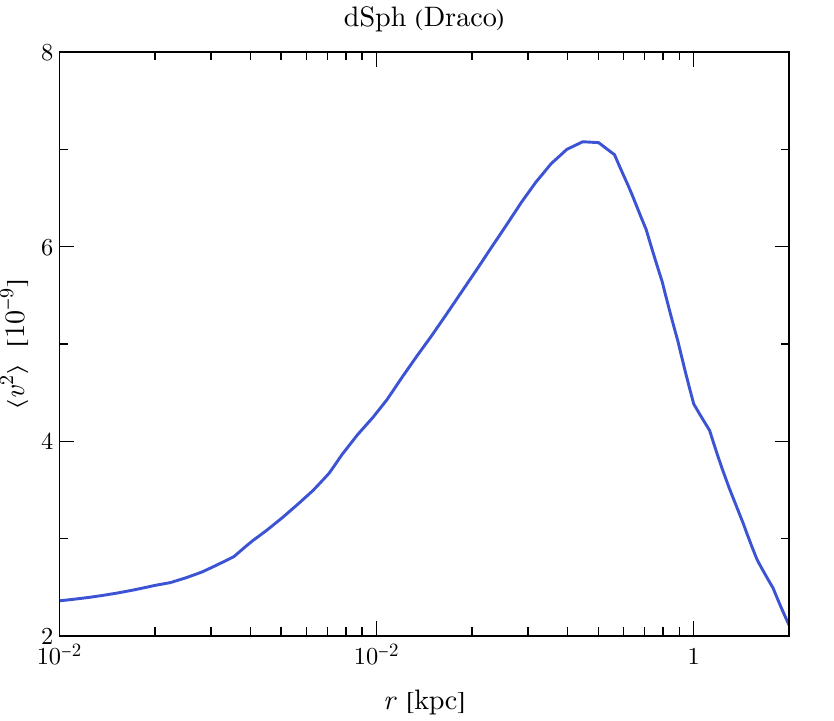}
\caption{The mean square velocity of the dark matter in the halo of the Milky Way (left panel) and in the Draco dwarf galaxy (right panel), each as a function of galactocentric radius (note the different ranges shown on the axes of the two panels).
}
\label{fig:velocityprofile}
\end{figure}

In this section, we apply the formalism presented in the previous section to the case of realistic DM density and velocity profiles, in order to calculate the abundances and distributions of $\x_2$ particles in the Milky Way and in dwarf galaxies. For the Milky Way, we adopt an NFW density profile~\cite{Navarro:1996gj}, 
\be
\rho_\x^\text{NFW} (r) = \frac{\rho_s}{(r / r_s) \, (1 + r / r_s)^2}
~,
\ee
where $r_s \approx 20 \ \kpc$ is the scale radius and $\rho_s \approx 0.35 \ \GeV/{\rm cm}^3$ is normalized such that the local DM density is $\rho \approx 0.4 \ \GeV / \cm^3$ at $r \approx 8.5 \ \kpc$. For dwarf galaxies, we instead implement a cored and tidally stripped NFW profile,
\be
\rho_\x^\text{cNFWt} (r) =
\begin{cases}
\rho_\x^\text{cNFW} (r) & r \leq r_t
\\
\rho_\x^\text{cNFW} (r) ~ (r/r_t)^{-\delta} & r > r_t
~,
\end{cases}
\ee
where $r_t$ is the tidal radius. This is written in terms of the unstripped cored profile, which is given by
\be
\rho_\x^\text{cNFW} (r) = f_c^n \, \rho_\x^\text{NFW}(r) + \frac{n \, f_c^{n-1} \, (1-f_c^2)}{4 \pi r^2 \, r_c} \, M_\text{NFW}(r)
~,
\ee
where $f_c = \tanh{(r/r_c)}$, $r_c$ is the core radius, $n$ and $\delta$ are dimensionless positive numbers, and 
\be
M_\text{NFW} (r) = 4 \pi \, \rho_s \, r_s^3 \, \bigg[ \ln{\big( 1 + r / r_s \big)} - \frac{r}{r + r_s} \bigg]
\ee
is the enclosed NFW mass. We adopt values for the parameters $\rho_s$, $r_s$, $r_c$, $r_t$, $n$, and $\delta$ for specific dwarf galaxies, as provided in Ref.~\cite{DiMauro:2022hue}.

As shown in, e.g., Ref.~\cite{1987book}, the DM velocity distribution, $f(v)$, can be derived from the DM density profile using the Eddington inversion formula. Up to a normalization constant, this yields 
\be
f (v) \propto 
\int^0_{\eps(v)} \frac{d \Phi}{\sqrt{\Phi - \eps(v)}} \, \frac{d^2 \rho_\x}{d \Phi^2}
~,
\ee
where $\eps(v) \equiv v^2 / 2 + \Phi  < 0$ is the mass-normalized energy, and $\Phi (r) < 0$ is the gravitational potential. We evaluate the gravitational potential of each dwarf galaxy using the DM density profiles discussed above. For the Milky Way, we follow Ref.~\cite{Boddy:2018ike} and adopt a spherically-symmetric approximate form of the Galactic potential that includes contributions from a standard NFW DM profile, as well as from the baryonic disk and bulge. For a given $\Phi (r)$, we can invert this to calculate $r(\Phi)$, and hence obtain $\rho_\x(\Phi)$ from $\rho_\x (r)$. This can then be used in the integral above to calculate the velocity distribution, $f(v)$. Our results are shown in \Fig{velocityprofile}, where we plot the mean of the square of the DM's velocity as a function of galactocentric radius for both the Milky Way and the Draco dwarf galaxy. 

\begin{figure}[t!]
\centering
\includegraphics[width=0.6\columnwidth]{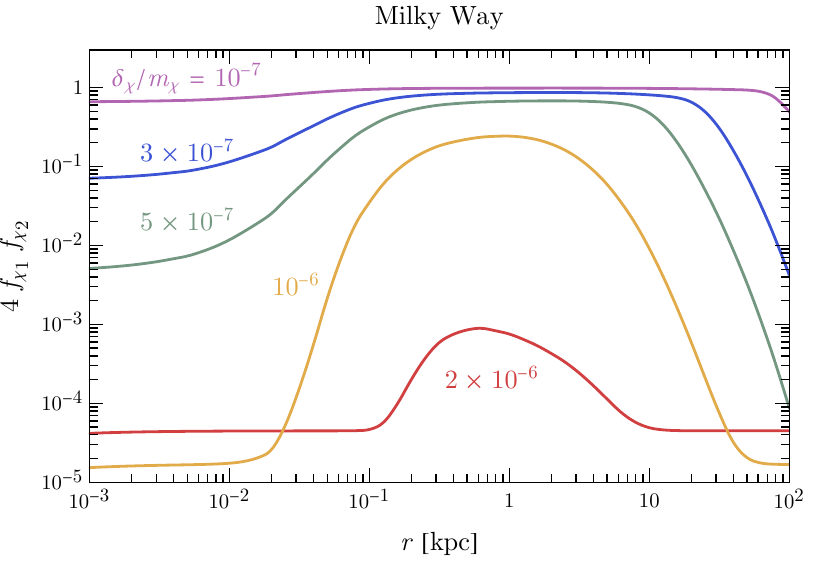}
\caption{The quantity $4 f_{\x_1} f_{\x_2}$ for the Milky Way, where $f_{\x_i} \equiv n_{\x_i}/(n_{\x_1}+n_{\x_2})$, as a function of galactocentric radius, for several values of the dark matter's fractional mass splitting, $\delta_\x/m_\x$, and for the model parameters given in Table~\ref{tab:parm}. We have further adopted $y_\x^\p = y_\x^{\p\,{\rm max}}$, where $y_\x^{\p\,{\rm max}}$ is the value of this coupling for which the processes $\x_1 \x_1 \to \varphi \varphi$ and $\x_2 \x_2 \to \varphi \varphi$ contribute 10\% of the total annihilation rate at the time of thermal freeze out.}
\label{fig:f1f2profb}
\end{figure}

We use these DM density and velocity profiles to solve the Boltzmann equations of \Sec{upscatter} and determine the fractional abundance of $\x_2$ particles, $f_{\x_2}$, as a function of galactocentric radius. In doing so, we approximate the effective DM temperature as $T_\x \approx m_\x \, \langle v^2 \rangle / 3$, where $\langle v^2 \rangle$ is determined from the velocity distribution discussed above. We have also made the approximation that each $\x_2$ remains at a fixed radius, $r$, such that the Boltzmann equation at each value of $r$ can be evaluated independently. In \Fig{f1f2profb}, we show the resulting radial profile of $4 f_{\x_1} f_{\x_2}$ in the Milky Way, for various values of the fractional mass splitting, $\delta_\x/m_\x$, and a representative choice of the other model parameters. Due to the fact that the DM velocity peaks at intermediate radii (left panel of \Fig{velocityprofile}), the $\x_2$ density is peaked at intermediate radii as well.

For standard DM annihilations, the gamma-ray flux is determined by the $J$-factor,  
\be
J_0 = \int d \Omega \, d \ell ~ \rho_\x (r)^2
~,
\ee
where the solid angle, $\Omega$, and radial coordinate, $\ell$, defined with respect to a terrestrial observer, are related to galactocentric coordinates by $r^2 = r_\odot^2 + \ell^2 - 2 r_\odot \, \ell \, \cos{\theta}$ and $r_\odot \approx 8.5 \ \kpc$. For the Galactic Center, we average over the inner $20^{\degree}$, whereas for a typical dwarf galaxy we average over $1^{\degree}$. Since the two DM states, $\x_1$ and $\x_2$, have distinct density profiles in this model, we define the analogous $J$-factor for our analysis,
\be
J_\x = 4 \int d \Omega \, d \ell ~ \rho_\x(r)^2 \, f_{\x_1} (r) \, f_{\x_2} (r)
~,
\ee
where, once again, we include the factor of four in order to facilitate a straightforward comparison with conventional Dirac DM. 

\begin{figure}[t!]
\centering
\includegraphics[width=0.6\columnwidth]{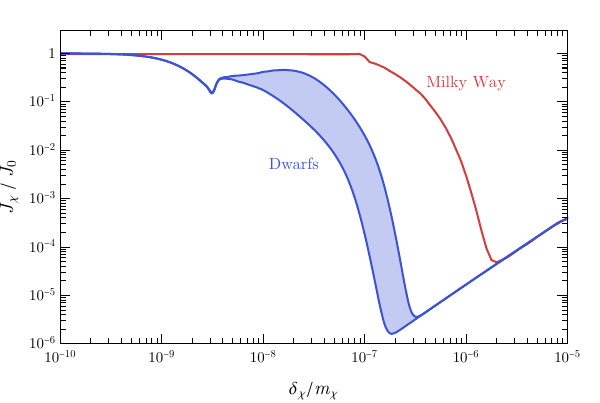}
\caption{The suppression in the effective dark matter annihilation rate, $J_\x / J_0$, for the Milky Way and dwarf galaxies, as a function of the fractional mass splitting. For dwarf galaxies, we adopt a density and velocity profile of a Draco-like dwarf galaxy, located at a distance between $(50-200) \ \kpc$ from the Galactic Center (closer/further distances correspond to larger/smaller values of $J_\x / J_0$), corresponding to the shaded blue region. Here, we have adopted the model parameters given in Table~\ref{tab:parm}, as well as $y_\x^\p = y_\x^{\p \, \text{max}}$.}
\label{fig:Jfactor}
\end{figure}

In \Fig{Jfactor}, we show the suppression in the effective $J$-factor, $J_\x / J_0$, for the Milky Way and dwarf galaxies, as a function of the fractional mass splitting and for a representative choice of model parameters. For dwarf galaxies, we adopt the density and velocity profiles of a Draco-like galaxy. The shaded blue region corresponds to the range of distances between typical dwarf galaxies and the Galactic Center. In generating this figure, we have assumed that virialization at a given radius occurs on a timescale comparable to the period of a circular orbit at that radius. If DM upscattering at a given radius occurs on a shorter timescale, we assume that the regenerated $\x_2$ particles remain at that radius, as in \Fig{f1f2profb}. If upscattering instead occurs on a longer timescale, we assume that both the $\x_1$ and $\x_2$ populations ``revirialize," readjusting to the profile that existed prior to upscattering, normalized to conserve particle number. See Appendix~\ref{sec:sub} for further details.

The quantity $J_\x / J_0$ shown in \Fig{Jfactor} corresponds to the relative suppression in the gamma-ray flux from DM coannihilation, $\x_1 \x_2 \to f \bar{f}$, compared to standard DM scenarios of the same mass and cross section. Hence, $J_\x / J_0 = 1$ corresponds to a coannihilation gamma-ray flux comparable to Dirac DM with a velocity-independent annihilation cross section of $\langle \sigma v \rangle \approx 4 \times 10^{-26} \, \cmps$.  However, we note that for extremely small values of $J_\x / J_0$, other higher-order processes may dominate over coannihilations. For instance, one could consider the annihilation of two $\x_1$ particles through two off-shell pseudoscalars to four SM fermions, or through a loop of pseudoscalars to a pair of SM fermions. In either case, a simple estimate predicts that such annihilation rates are at least $\sim 10^{-12}$ times smaller than the rate of coannihilation, $\sigma v_{\x_1 \x_2 \to f \bar{f}}$. Since the smallest suppression to the coannihilation rate in \Fig{Jfactor} is $J_\x / J_0 \gtrsim 10^{-6}$, we can safely neglect such higher-order annihilation processes.

\section{Summary and Discussion}
\label{sec:conclusion}

The brightest gamma-ray signal from dark matter annihilation is expected to come from the region surrounding the Galactic Center. That region, however, also produces a sizable flux of gamma rays through a variety of astrophysical processes and sources, and this could make it difficult to robustly identify any would-be signals of annihilating dark matter. On the other hand, the Milky Way's dwarf spheroidal galaxies are highly dark matter-dominated systems which produce little in the way of astrophysical backgrounds. This makes dwarf galaxies valuable as targets of indirect dark matter searches, despite the fact that they are expected to produce much lower gamma-ray fluxes from dark matter annihilation.

The Galactic Center Gamma-Ray Excess is statistically significant and has a spectral shape, angular distribution, and overall normalization which are consistent with those expected from annihilating dark matter in the form of a $\sim 50 \ \GeV$ thermal relic. It has alternatively been suggested that these gamma rays could arise from a large population of faint and centrally-concentrated millisecond pulsars. Dwarf galaxies could be instrumental in settling this debate. 

In this paper, we have presented a dark matter model in which the connection between the gamma-ray flux from the Galactic Center and from dwarf galaxies is broken. In particular, the dark matter annihilation rate in dwarf galaxies can be highly suppressed in this model relative to that which would be naively expected from the annihilation rate in the Galactic Center. Gamma-ray signals are generated in this model through the coannihilations of two dark matter states with a small mass splitting. Dark matter particles can only be upscattered into their excited state, $\x_1 \x_1 \rightarrow \x_2 \x_2$, if they have enough kinetic energy to overcome this mass splitting. This allows a large $\x_2$ abundance to be generated in the Galactic Halo, but not in the lower-velocity environments of dwarf galaxies. This, in turn, can lead to a detectable gamma-ray signal from the Galactic Center, while suppressing any corresponding signal from dwarf galaxies. 

In addition to the specific model presented in this study, there are other examples from which similar phenomenology could arise.  For example, one could consider an alternative model whose Lagrangian contains
\be
\Lag \supset \lambda \, \varphi \, \bar{\x}_1 \x_2  + e^\p \Ap_\mu \, \bar{\x}_2 \gamma^\mu \x_2
~,
\ee
where $\varphi$ and $\Ap$ are scalar and vector mediators with couplings, $\lambda$ and $e^\prime$. In this variation, both mediators 
are light, and only $\Ap$ couples (very feebly) to the Standard Model. In this scenario, $\x_1 \x_1 \to \x_2 \x_2$ upscattering proceeds through $\varphi$ exchange, and $\x_1 \x_1 \to \varphi \varphi$ annihilation does not produce any visible signals. A gamma-ray signal is generated in this model only through the processes $\x_1 \x_2 \to \varphi \Ap$ or $\x_2 \x_2 \to \Ap \Ap$, followed by $\Ap \to f \bar{f}$ decays to Standard Model final states. We leave further investigation of such model variations to future work.

\section*{Acknowledgments}

This manuscript has been authored in part by Fermi Forward Discovery Group, LLC under Contract No. 89243024CSC000002 with the U.S. Department of Energy, Office of Science, Office of High Energy Physics.  DH is supported by the Office of the Vice Chancellor for Research at the University of Wisconsin–Madison with funding from the Wisconsin Alumni Research Foundation. This research was authored in part
by support from the Kavli Institute for Cosmological Physics at the University of Chicago through an
endowment from the Kavli Foundation and its founder
Fred Kavli
and also by
grant NSF PHY-2309135 to the Kavli Institute for Theoretical Physics (KITP) 

\begin{appendix}
\section{Dark Matter Temperature Evolution}
\label{sec:tempevol}

Here, we discuss the evolution of the DM temperature, $T_\x$. In the very early universe, the temperature of the DM is maintained in equilibrium with that of the SM bath, $T$. This continues until the rate of fractional energy transfer, $\Gamma^E$, from DM scattering with SM fermions drops below twice the rate of Hubble expansion~\cite{Gondolo:2012vh}:
\be
2 H \gtrsim \Gamma^E_{\x_2 f \to \x_1 f} \approx
\frac{y_\x^2 \, y_f^2 \, T^4}{m_\x^3 \, m_a^4}
\times
\begin{cases}
(127 \pi^5 / 90) \, T^4
& (m_f \ll T \ll m_\x)
\\
(128 / \pi^3) \, m_f^4 \, e^{-m_f / T}
& (T \ll m_f \ll m_\x)
~.
\end{cases}
\ee
We denote the temperature at which this departure from kinetic equilibrium occurs by $T_{\x f}^\text{KD}$. 

After $\x$ kinetically decouples from the SM, the $\x$ and $\varphi$ populations remain kinetically coupled by the scattering process, $\x_1 \varphi \to \x_1 \varphi$. During this time (at $T_\x < T^{\rm KD}_{\x f}$), the DM temperature  evolves according to $T_\x \approx \big[ g_{* S} (T) / g_{* S} (T_{\x f}^\text{KD}) \big]^{1/3} \, T$, which follows from entropy conservation. The condition for kinetic equilibrium to be maintained between the $\x_1$ and $\varphi$ populations is given by
\be
2H \lesssim \Gamma^E_{\x_1 \varphi \to \x_1 \varphi} \approx \frac{y_\x^{\p \, 4} \, T_\x^4}{\pi^3 \, m_\x^3}
\times
\begin{cases}
(180 \, \zeta(5) - \pi^4) / 540
& (m_\varphi \ll T_\x \ll m_\x)
\\
(8 / 3) \, e^{- m_\varphi / T_\x}
& (T_\x \ll m_\varphi \ll m_\x)
~.
\end{cases}
\ee
The SM temperature at which this condition ceases to be satisfied is denoted by $T_{\x \varphi}^\text{KD}$.

Applying entropy conservation, the full DM temperature evolution is given by
\be
T_\x \approx
\begin{dcases}
T & (T \geq T_{\x f}^\text{KD})
\\ 
T_{\x f}^\text{KD} ~ \bigg(\frac{a (T_{\x f}^\text{KD})}{a(T)}\bigg) & (T_{\x \varphi}^\text{KD} \leq T < T_{\x f}^\text{KD})
\\ 
T_{\x f}^\text{KD} ~ \bigg(\frac{a (T_{\x f}^\text{KD})}{a(T_{\x \varphi}^\text{KD})}\bigg) ~  \bigg( \frac{a (T_{\x \varphi}^\text{KD})}{a(T)} \bigg)^2  & (T \leq T_{\x \varphi}^\text{KD} < T_{\x f}^\text{KD})
\\ 
T_{\x f}^\text{KD} ~  \bigg( \frac{a (T_{\x f}^\text{KD})}{a(T)} \bigg)^2  & (T \leq T_{\x f}^\text{KD} <  T_{\x \varphi}^\text{KD})
~.
\end{dcases}
\ee
In our analysis, we implement this full temperature evolution of the DM. For the canonical choice of model parameters adopted throughout this work, we find that for $y_\x^\p \gg 10^{-2} \times y_\x^{\p \, \text{max}}$, $\x - \varphi$ kinetic decoupling typically occurs well after $\x - f$ kinetic decoupling.

\section{Downscattering}
\label{sec:downscattering}

For $m_{\varphi} \ll T_\x \ll m_\x$ and $m_{\varphi} \ll \delta_\x m_\x^{1/2}/T^{1/2}_\x$, the thermally-averaged cross section for the process $\x_2 \x_2 \to \x_1 \x_1$ takes a simple analytic form. In particular, in the limit that $y_\x^\p \ll v^{1/2}$, the rate is well-approximated by the tree-level exchange of $\varphi$ (Born approximation), which yields
\be
\label{eq:down}
\lim_{y_\x^{\p \, 2} \ll v} \, \langle \sigma v\rangle_{\x_2 \x_2 \to \x_1 \x_1}
\approx
\frac{y_\x^{\p \, 4}}{4 \pi^{3/2}} \, \frac{T_\x^{1/2}}{\delta_\x^2 \, m^{1/2}_\x }
~ .
\ee
In the opposite limit, where $y_\x^\p \gg v^{1/2}$, which applies to most of the model space investigated in this work, this rate is Sommerfeld enhanced by the long-ranged $\varphi$ potential. To incorporate this effect, we assume that the DM mass splitting is small and adopt the semi-analytic results presented in Refs.~\cite{Cirelli:2016rnw,Petraki:2015hla} (Sommerfeld enhancements in the presence of mass splittings have also been studied previously in Refs.~\cite{Slatyer:2009vg,Finkbeiner:2007kk,Chen:2009dm}). For fixed relative velocity, this corresponds to an enhancement of roughly $\sim  y_\x^{\p\,2} / 2 v$, which can be much larger than unity when $\x_{1,2}$ are highly non-relativistic. In terms of the DM temperature, we take this to be $y_\x^{\p \, 2} \langle 1 / 2 v \rangle \approx y_\x^{\p \, 2} \sqrt{m_\x / 2 \pi T_\x}$. Including this factor in \Eq{down} then gives the result shown earlier in \Eq{sigmavdown}.

Note that \Eqs{down}{sigmavdown} are enhanced in the limit of a small mass splitting, $\delta_\x \ll m_\x$. This is regulated by the mass of $\varphi$, such that the cross section becomes independent of the mass splitting and instead scales as $\propto m_\varphi^{-2}$ when $\delta_\x \lesssim m_{\varphi} \, v$, corresponding to a mediator heavier than the minimum required momentum transfer. This feature is not present in most rates related to the astrophysical signatures of self-interacting DM (SIDM). This is because it is the amount of momentum/energy that is transferred in those collisions that is relevant in those studies. In particular, the relevant quantity for the momentum-exchange of indistinguishable DM particles is instead the so-called viscosity cross section~\cite{Knapen:2017xzo,Tulin:2013teo}, which penalizes scatters which exchange a small amount of momentum. To illustrate this, we can use the same amplitude for $\x_2 \x_2 \to \x_1 \x_1$ to determine the viscosity cross section. For instance, setting $\delta_\x = 0$ and taking the small $m_\varphi$ limit, we find
\be
\label{eq:viscosity}
\sigma^\text{visc}_{\x_2 \x_2 \to \x_1 \x_1} \approx \frac{2 \, y_\x^{\p \, 4}}{\pi \, m_\x^2 \, v^4} ~ \ln \bigg( \frac{m_\x \, v }{m_\varphi} \bigg)\,,
\ee
which, unlike \Eq{down}, is only logarithmically enhanced by the light mediator. As we are interested here in particle conversion rather than momentum-exchange, we use the cross section given in \Eq{sigmavdown}, which does not penalize forward or backward scattering.

The $\x_2$ population can also downscatter off of SM fermions through the process $\x_2 f \to \x_1 f$, which involves the exchange of the $a$ mediator. Compared to $\x_2 \x_2 \to \x_1 \x_1$, this process is suppressed due to $m_a \gg m_\varphi$, but is enhanced by the larger density of fermion targets at low temperatures. Taking $m_a \gg m_\x \gg m_f \gg  \delta_\x$, the thermally-averaged rate for this process is given by
\begin{align}
\Gamma_{\x_2 f \to \x_1 f} &\approx -\frac{32 \, y_\x^2 \, y_f^2 \, T_f^4}{\pi^3 \, m_\x^2 \, m_a^4} ~ \Bigg \{ m_f^3 \, \text{Li}_4 \big( -e^{-m_f / T_f} \big) +3 T_f \bigg[ 2 m_f^2 \, \text{Li}_5 \big( -e^{-m_f / T_f}\big)
\nl
& \qquad \qquad \qquad  \quad +5 T_f  \, \Big(m_f \, \text{Li}_6 \big( -e^{-m_f / T_f} \big)+T_f \, \text{Li}_7 \big(-e^{-m_f / T_f}\big)\Big)\bigg] \Bigg\}
\nl
& \approx
\frac{y_\x^2 \, y_f^2 \, T_f^4}{\pi^3 \, m_\x^2 \, m_a^4}
\times
\begin{cases}
945  \, \zeta(7) \, T_f^3 / 2 & (m_f \ll T_f \ll m_\x)
\\
32 \, m_f^3 \, e^{-m_f / T_f} & (T_f \ll m_f \ll m_\x)
~,
\end{cases}
\end{align}
where $T_f$ is the temperature of $f$. In the parameter space of interest, we find that the process $\x_2 f \leftrightarrow \x_1 f$ becomes inefficient at DM temperatures $T_\x \gg \delta_\x$, well before the rate of $\x_2 \x_2 \leftrightarrow \x_1 \x_1$ falls below that of Hubble expansion.

\section{Upscattering and Dark Matter Substructure}
\label{sec:sub}

We noted in \Sec{upscatter} that the simple Boltzmann equation of \Eq{Boltzmann1} describes the evolution of an isolated DM system, and is therefore not a good approximation for dwarf galaxies, which reside within the larger halo of the Milky Way. As a result, even for mass splittings well above the characteristic temperature of the dwarf galaxy, $\delta_\x / m_\x \gtrsim 10^{-9}$, slowly moving $\x_1$ particles in the dwarf galaxy can upscatter with the higher-velocity Galactic $\x_1$ population into a pair of $\x_2$ particles. For one of the resulting $\x_2$ particles to remain bound to the dwarf galaxy, the imparted velocity must not exceed the dwarf galaxy's escape velocity, which requires $2 \delta_\x / (m_\x v_\text{MW}) \lesssim v_\text{dSph}$, or equivalently, $\delta_\x / m_\x \lesssim 10^{-8}$. To account for the effect of such processes on the dwarf galaxy's fractional DM abundances, we use the following modified form of the Boltzmann equation:
\begin{align}
\label{eq:Boltzmann2}
\dot{f}_{\x_2}^\text{dSph} &= \Big( \big( f_{\x_1}^\text{dSph} \big)^2 \, e^{- 2 \delta_\x / T_\x^\text{dSph}} - \big( f_{\x_2}^\text{dSph} \big)^2 \Big) \, \frac{\rho_\x^\text{dSph}}{m_\x} \, \sigv_{\x_2 \x_2 \to \x_1 \x_1}^\text{dSph}
\nl
&+ \Big(  f_{\x_1}^\text{dSph} \, f_{\x_1}^\text{MW} \, e^{- 2 \delta_\x / T_\x^\text{MW}} - f_{\x_2}^\text{dSph} \, f_{\x_2}^\text{MW} \Big) \, \frac{\rho_\x^\text{MW}}{m_\x} \, \sigv_{\x_2 \x_2 \to \x_1 \x_1}^\text{MW} \, e^{-2 \delta_\x / \big(3 \sqrt{T_\x^\text{MW} \, T_\x^\text{dSph}}\big)}
\nl
\dot{f}_{\x_1}^\text{dSph} &= - \dot{f}_{\x_2}^\text{dSph}
~,
\end{align}
where the superscripts denote whether these quantities refer to the dwarf galaxy or to the Milky Way halo (with the latter evaluated at the position of the dwarf galaxy). The first line of \Eq{Boltzmann2} corresponds to upscattering solely among DM particles belonging to the dwarf galaxy, $\x_1^\text{dSph} \x_1^\text{dSph} \to \x_2^\text{dSph} \x_2^\text{dSph}$, whereas the second line accounts for scattering between Galactic and dwarf galaxy particles, $\x_1^\text{MW} \x_1^\text{dSph} \to \x_2^\text{MW} \x_2^\text{dSph}$. We have added the final exponential factor in the second line to account for the fact that one of the $\x_2$ particles produced in interactions between the dwarf galaxy and the halo of the Milky Way does not remain bound to the former unless $2 \delta_\x / (m_\x v_\text{MW}) \lesssim v_\text{dSph}$, as discussed above. 

Note that the evolution of $f_{\x_{1,2}}^\text{dSph}$ in \Eq{Boltzmann2} depends directly on $f_{\x_{1,2}}^\text{MW}$, which in turn depends on the degree to which these two systems are coupled. In particular, if
\be
\label{eq:MWcoupling}
f_{\x_1}^\text{MW} \, \frac{\rho_\x^\text{MW}}{m_\x} \, e^{- 2 \delta_\x / T_\x^\text{MW}} \, \sigv_{\x_2 \x_2 \to \x_1 \x_1}^\text{MW} \gg \frac{v_\text{orbit}^\text{dSph}}{r_\text{dSph}}
~,
\ee
then a given $\x_1$ particle in the halo of the Milky Way will strongly couple to the dwarf galaxy, where $v_\text{orbit}^\text{dSph}$ is the orbital speed of the dwarf galaxy and $r_\text{dSph}$ is its tidal radius. In the case that \Eq{MWcoupling} holds, we supplement \Eq{Boltzmann2} with the analogous equation for the halo of the Milky Way. Alternatively, if \Eq{MWcoupling} does not hold, we determine the Milky Way fractional abundances using \Eq{Boltzmann1}, and use these as an input to \Eq{Boltzmann2}. For most of the mass splittings considered in this work, \Eq{MWcoupling} does not hold, in which case it is simple to determine the point at which $f_{\x_2}^\text{dSph}$ stops evolving and enters detailed balance. In particular, setting \Eq{Boltzmann2} to zero and assuming that $v_\text{dSph}^2 \lesssim \delta_\x / m_\x \lesssim v_\text{dSph} \, v_\text{MW}$, this occurs for $f_{\x_2}^\text{dSph} \sim \sqrt{\rho_\x^\text{MW} / \rho_\x^\text{dSph}} \ll 1$. For a typical dwarf galaxy, which resides $\sim 100 \ \kpc$ from the Galactic Center, $\rho_\x^\text{MW} \sim \text{few} \times 10^{-3} \ \GeV / \cm^3$ and $\rho_\x^\text{dSph} \sim 1 \ \GeV / \cm^3$, thus implying that $f_{\x_2}^\text{dSph} \sim \text{few} \times 10^{-2}$. This is evident in the results shown in the left panel of \Fig{f1f2} for $10^{-9} \lesssim \delta_\x / m_\x \lesssim 10^{-8}$.

\begin{figure}[t!]
\centering
\includegraphics[width=0.6\columnwidth]{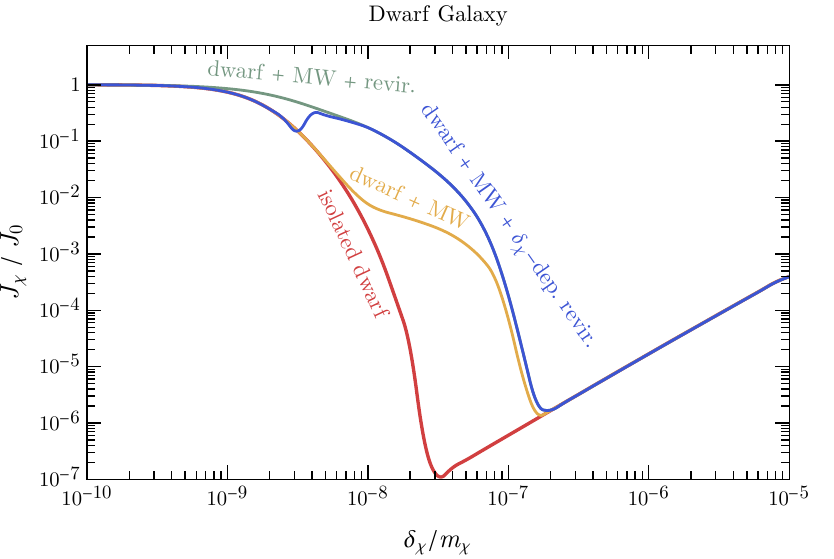}
\caption{As in \Fig{Jfactor}, but showing the effect of various model assumptions in the calculation of the $J$-factor for dwarf galaxies. See the discussion in Appendix~\ref{sec:sub} for additional details.}
\label{fig:JfactorDwarf}
\end{figure}

To more clearly show the impact of the Milky Way's halo on the DM density of a typical dwarf galaxy, we performed a series of of calculations analogous to those shown in \Fig{Jfactor}, but under different assumptions. For the contour labeled ``isolated dwarf" in \Fig{JfactorDwarf}, we have neglected the effect of the Milky Way's halo on the dwarf galaxy and assumed that the $\x_2$ population remains unvirialized. For the line labeled ``dwarf $+$ MW," we incorporated the effect of the Milky Way, assuming that the dwarf is at a distance of $200 \ \kpc$ from the Galactic Center. For ``dwarf $+$ MW $+$ revir.," we took the $\x_2$ population to fully revirialize throughout the dwarf galaxy. Finally, for ``dwarf $+$ MW $+$ $\delta_\x$-dep. revir.," we incorporated the ``revirialization" criteria discussed near the end of \Sec{IDD}. The last result listed here corresponds to that shown for dwarf galaxies in \Fig{Jfactor}. From this, we see that the effects of revirialization and the Milky Way significantly increase the effective $J$-factor of the dwarf galaxy.

\section{Decays of the Heavy State}
\label{sec:decays}

Even if DM upscattering is very efficient in the halo of the Milky Way, the $\x_2$ must be very long-lived in order to produce a detectable gamma-ray signal through coannihilations with $\x_1$ at late times. For $\delta_\x < m_{\varphi}$, decays of the form $\x_2 \to \x_1 \, \varphi$ will be kinematically forbidden.
Thus, the decays of $\x_2$ can only proceed through an off-shell $a$ into SM fermions or photons. The rate for the former process in the $\delta_\x , m_f \ll m_\x$ limit is given by
\be
\Gamma_{\x_2 \to \x_1 f \bar{f}} \approx \frac{y_\x^2 \, y_f^2 \,  \delta_\x^7}{560 \pi^3 \, m_a^4 \, m_\x^2}
~.
\ee
This corresponds to a lifetime that is longer than the age of the universe if
\be
\frac{\delta_\x}{m_\x} \lesssim 5 \times 10^{-6} \, \bigg( \frac{1}{y_\x \, y_f} \bigg)^{2/7} \bigg( \frac{50 \ \GeV}{m_\x} \bigg)^{1/7} \, \bigg( \frac{m_a / m_\x}{3} \bigg)^{4/7}
~.
\ee
Note that this decay rate is proportional to $\delta_\x^7$, unlike the $\delta_\x^5$ scaling that is found in the case of a scalar or vector mediator~\cite{Izaguirre:2015zva,Berlin:2018jbm}. This arises due to the difference of the pseudoscalar interaction in the non-relativistic limit. 

Note that these decays will be irrelevant in our parameter space of interest, which features mass splittings that are too small to produce final state electrons. Decays of the form $\x_2 \to \x_1 \g \g$, however, can take place through a loop of quarks or charged leptons, arising from the effective operator,
\be
\Lag \supset \frac{g_{a \g \g}}{4} \, a \, F^{\mn} \tilde{F}_{\mn}
~,
\ee
where $\tilde{F}_{\mn} = \eps_{\mn \rho \sigma} \, F^{\rho \sigma} / 2$. The form of $g_{a \g\g}$ can be identified through the chiral anomaly. In particular, a chiral rotation of a fermion, $f$, can shuffle between two forms of the interaction,
\be
g_a \, \partial_\mu a \, \bar{f} \g^\mu \g^5 f \leftrightarrow 
- 2 m_f \, g_a \, a \, \bar{f} i \g^5 f - \frac{N_f \, q_f^2 \, \alpha_\text{em} \, g_a}{2 \pi} \, a \, F^{\mn} \tilde{F}_{\mn}
~, 
\ee
where $N_f$ is the number of fermion species and $q_f$ is its electromagnetic charge. Setting $g_a = - y_f / 2 m_f$ yields a second form of the Lagrangian,
\be
\Lag \supset y_f \, a \, \bar{f} i \g^5 f + \frac{N_f \, q_f^2 \, \alpha_\text{em} \, y_f}{4 \pi \, m_f} \, a \, F^{\mn} \tilde{F}_{\mn}
~.
\ee
The second term in this expression is effectively the radiative contribution from the first term. For the interaction in question, we have $y_f = \theta \,  \tan{\beta} \, m_f / v_h$ for loops of down-type quarks and charged leptons, yielding
\begin{align}
g_{a \g \g} &= \frac{\alpha_\text{em} \, \theta \, \tan{\beta}}{\pi \, v_h} \, \sum_f N_f \, q_f^2 = \frac{4 \, \alpha_\text{em} \, \theta \, \tan{\beta}}{\pi \, v_h}~, \nonumber 
\end{align}
where we summed over species of quarks and charged leptons. The resulting decay rate is then given by
\be
\Gamma_{\x_2 \to \x_1 \g \g} \approx \frac{\alpha_\text{em}^2}{630 \, \pi^5} \, \frac{(y_\x \, \theta \tan{\beta})^2 \, \delta_\x^9}{v_h^2 \, m_a^4 \, m_\x^2}
~.
\ee
For the parameter space of interest here, the corresponding lifetime is more than $10^{20}$ times the age of the universe, well beyond the reach of searches for DM decay products.

\end{appendix}

\bibliographystyle{utphys3}
\bibliography{biblio}

\end{document}